\def\L {{\cal L}}
\def\p {\partial}
\def\f {\phi^{ija\{\nu\}}_{\{\mu\}}}
\def\fu {\phi^{ija\{\nu\}_1\beta \{\nu\}_2}_{\{\mu\}}}
\def\fd {\phi^{ija\{\nu\}}_{\{\mu\}_1 \alpha \{\mu\}_2}}
\def\a {\alpha}
\def\b {\beta}
\def\r {\rho}
\def\d {\delta}
\def\D {{\cal D}}
\def\M {{\cal M}}

\documentstyle[preprint,eqsecnum,aps]{revtex}
\begin{document}
\draft
\preprint{UCSBTH-94-45, gr-qc/9411067}
\title{Diffeomorphism Invariant Actions for Partial Systems}
\author{Donald Marolf}
\address{Physics Department, The University of California,
Santa Barbara, California 93106} \date{November, 1994}
\maketitle

\begin{abstract}
 Local action principles on a manifold $\M$
 are invariant (if at all) only under
 diffeomorphisms that preserve the boundary of $\M$.
 Suppose, however, that we wish to study only part
 of a system described by such a principle; namely, the part
 that lies in a bounded
 region $R$ of spacetime where $R$ is specified in some
 diffeomorphism invariant manner.  In this case, a description of
 the physics within $R$ should be invariant under {\it all} diffeomorphisms
 regardless of whether they preserve the boundary of this region.
 The following letter shows that physics in such a
 region can be described by
 an action principle that $i$) is invariant under both diffeomorphisms
 which preserve the boundary of $R$ and those that do not,
 $ii$) leaves the dynamics of the part of the system {\it outside} the
 region $R$ completely undetermined, and $iii$) can be constructed
without first solving the original equations of motion.

\end{abstract}

\section{Introduction}

Action principles for systems with boundaries are a fundamental tool
in studying quantum effects in diffeomorphism invariant systems.
They have been used to derive black hole entropy in the
semiclassical approximation
\cite{HI} and to discuss the pair
creation and annihilation of magnetically charged black holes
\cite{pairs}.  Recently, Carlip \cite{steve} has even suggested
that the entropy of the 2+1 dimensional
Ban\~ados, Teitelboim, and Zanelli black hole \cite{BTZ} can be
derived by counting the states produced by the degrees of freedom
that arise through a ``restriction of the diffeomorphism invariance''
by the presence of a boundary; in that case, the horizon of a
black hole.  In addition, asymptotically flat and other
noncompact spacetimes may be described as systems with a boundary
\cite{roger}, and spacelike singularities form a natural
past or future boundary for many classical solutions of
general relativity and low energy string theory (see, e.g., \cite{sing}).
Finally, to render the action of {\it any} system finite, it is generally
necessary to consider the system only between two times or between
two spacelike hypersurfaces.
Thus, there is ample motivation to understand any subtleties that
arise in the use of variational principles for bounded generally covariant
systems.

The actions for such systems are typically of the local form
\begin{equation}
\label{s0}
S^{\M}_0 = \int_\M {\cal L} \ d^nx
\end{equation}
where $\M$ is an $n$-manifold and $\L$ is a scalar density on $\M$.
Such an action is invariant under any diffeomorphism
$\psi: \M\rightarrow \M$.  Note that such a map
induces a diffeomorphism of the boundary $\p\M$ of $\M$ as well.  If
the map did not preserve $\p\M$, it would
correspond to enlarging (or shrinking)
the system considered and would in general change the action.
While the addition of the proper boundary
term to \ref{s0} can enlarge the gauge invariance of $S_0^{\cal M}$
\cite{HTV,conf},
this will in general require the full solution of
the equations of motion.  As
a result, if the action $S_0^{\cal M}$
defines the notion of gauge equivalence,
an infinitesimal map $\delta \psi$ is a gauge transformation only if
it preserves $\p\M$.

Suppose now that $\M$ may be embedded in some larger manifold
$\M'$.  We would like to understand how the notions of gauge invariance
defined by $S_0^{\M}$ and $S_0^{\M'}$ relate.   Because
$S_0^{\M'}$ is invariant under diffeomorphisms that move the
boundary of $\M$, $S_0^{\M'}$ provides a larger set of infinitesimal
gauge transformations than does $S_0^{\M}$, even within the
image of $\M$.  Thus, the class of gauge invariants defined by
$S_0^{\M}$ is larger than that defined by $S_0^{\M'}$.  In particular,
if the theory contains a metric $g_{\mu \nu}$ and $\M$ is compact,
the quantity
\begin{equation}
\label{vol}
\int_{\M} \sqrt{-g} d^nx
\end{equation}
is invariant with respect to gauge transformations defined by
$S_0^{\M}$, but not those defined by $S_0^{\M'}$.

When the boundaries of $\M$ are not specified
by a physical condition, the action $S_0^{\M}$ and $S_0^{\M'}$
seem to describe quite different physics.
If, however, the above embedding is chosen in a field dependent but
diffeomorphism invariant manner (such as by mapping timelike
boundaries to sheets of steel and spacelike boundaries to
hypersurfaces defined by the reading of some clock), this
picture is physically reasonable as \ref{vol} may be interpreted as the
spacetime volume of the region bounded by the steel sheets for the
appropriate clock readings.  Such a quantity is gauge invariant as defined by
$S_0^{\M'}$ as well.  We would like to make this
connection explicit by describing the part of the system within
such boundaries in a way that
is invariant under all diffeomorphisms of $\M'$, even those that
move $\p \M$.

The purpose of this letter is to use this physical picture to
provide an action principle which achieves these goals {\it without}
first solving the equations of motion.  In
particular, given any action principle of the form \ref{s0} and
any (not necessarily local!) scalar field $f$, we show that
variation of the action
\begin{equation}
\label{newS}
S^{\M} = \int_{\M} \theta(f) {\cal L} \ d^nx
\end{equation}
where $\theta$ is the Heaviside step function,
yields the same Euler-Lagrange equations as $S_0^{\M}$ in the region where
$f>0$ {\it but provides no other restrictions on the dynamics}
when varied within a diffeomorphism invariant class of field
histories.  This property is nontrivial only on the surface $f=0$,
but, as should be expected, follows on this surface
only if the variations preserve appropriate
`boundary conditions' on the field histories.
Note that, provided $f < 0$ on the boundary of $\M$, the
action $S^{\cal M}$ is invariant under a larger class of gauge transformations
than $S_0^{\cal M}$.  Effectively, it {\it is} invariant under
transformations that
move the boundary of $\M$.  These results will be derived in the next
section.

\section{The Variational Principle for $S^{\cal M}$}

We now consider a general coordinate invariant
action principle of the form \ref{s0}
where the Lagrangian density ${\cal L}$ is a function of some
collection of fields and their first derivatives.  An action
of this form yields a well-defined variational principle whenever
all fields whose derivatives appear in ${\cal L}$ are
fixed on the boundary.  We refer to such fields as type I; fields
whose derivatives do not appear in ${\cal L}$ will be referred to
as type II.  However, for some first order systems such as spinor fields
or Chern-Simons fields \cite{ex1,ex2} it is only appropriate to specify
certain parts of these fields and to do so in a generally covariant
manner typically involves complicated nonlocal constructions\footnote{
Thanks to Steve Carlip for bringing this to the author's attention.}.
We do not treat such cases, but we expect that they may be
addressed by an analysis similar to
what follows.  We point out that our analysis {\it is}
appropriate to standard first order formulations of systems, like
gravity in any number of dimensions, which also have a second order
formulation.  In addition, although
not manifestly so, the usual action
\begin{equation}
{1 \over {16 \pi}} \int_{\M} \sqrt{-g} R + {1 \over {8\pi}} \int_{\p \M}
K
\end{equation}
for general relativity {\it is} of this form, as the boundary
term exactly cancels a total divergence which contains the higher
derivatives of $g_{\mu \nu}$.
More general variational principles, where
various momenta are fixed on the boundary, can be obtained from
actions of the above form by adding a total divergence to ${\cal L}$.  If the
Lagrangian is in first order form (see, for example, \cite{KK}),
the addition of such a divergence
leaves ${\cal L}$ a function only of the fields and their first
derivatives.  Thus, by passing through the first order
formulation, we see that actions of this type are
quite general.

Suppose that the Lagrangian depends on a set of
fields (labeled by $a$) of tensor type $i = (i_1,i_2)$ and
density weight $j$ which we write as $\phi^{ija\{\nu\}}_{\{\mu\}}$.
Here, $\{\nu\}$ and $\{\mu\}$ denote the appropriate collection
of abstract tensor indices as specified by the value of $i$.
${\cal L}$ will generally depend on fields of more than one
tensor type so that both $i$ and $j$ will take multiple values.
We employ the summation convention on the indices $i,j,a$ as
well as $\{\nu\}$ and $\{\mu\}$ so that the expression
\begin{equation}
j {{\partial {\cal L}} \over {\partial \f}}
\f \end{equation}
represents a sum over fully contracted fields of all tensor types
and all density weights with the contribution of each field
multiplied by its density weight.

It will be convenient to introduce an arbitrary background
spacetime connection $\Gamma^{\sigma}_{\alpha \beta}$ together
with the
associated covariant derivative operator ${\cal D}_{\alpha}$ and
to write ${\cal L}$ as a function of the
$\phi^{ijk\{\nu\}}_{\{\mu\}}$
and the ${\cal D}_{\alpha} \phi^{ijk\{\nu\}}_{\{\mu\}}$.  Since this
connection was introduced by hand, the fields and derivatives must
enter the Lagrangian in such a way that ${\cal L}$
is independent of $\Gamma^{\sigma}_{\alpha \beta}$.  We may thus
vary $\Gamma^{\sigma}_{\alpha \beta}$ as well in our action
principle and, while the resulting equation of motion will be
identically satisfied, this provides a convenient way
to keep track of the relationships that conspire to make
${\cal L}$ independent of $\Gamma^{\sigma}_{\alpha \beta}$ and thus
make $S_0^{\cal M}$ diffeomorphism invariant.

We will need the explicit form of the change of ${\cal L}$
under an infinitesimal coordinate transformation $x^\mu
\rightarrow x^\mu + \d x^\mu$:
\begin{eqnarray}
\label{dx}
\delta{\cal L} = &-& { {\partial {\cal L}} \over {\partial \f}}
[ \p_\a \f \ \delta x^\a + j \f \p_\a \  \delta x^\a
+ \sum_k \fd \ \p_{\mu_k} \d x^\a - \sum_k \fu \ \p_\b \d x^{\nu_k}]
\cr
&-& {{\p \L} \over {\p (\D_\r \f)}} [ \p_\a (\D_\r \f) \ \d x^\a
+ j \D_\r \f \ \p_\a  \d x^\a  + \D_\a \f \  \p_\r \d x^\a
\cr && \ \ \ \ \ \ \ \ \ \ + \sum_k \D_\r \fd \ \p_{\mu_k}   \d x^\a
- \sum_k \D_\r \fu \  \p_{\b}   \d x^{\nu_k}]
\end{eqnarray}
where the terms in square brackets are the explicit forms of
$\delta \f$ and $\delta \D_\r \f$ under a change of
coordinates.  The notation $\sum_k \fd \p_{\mu_k} \d x^\a$ represents
a sum whose $k$th term has the $k$th covariant index in the
set $\{\mu\}$ replaced by $\a$ and is contracted on that index
with $\p_{\mu_k}\d x^\a$,
where $\mu_k$ is just this missing index.  The corresponding notation
is used for the contravariant case as well.
Equation \ref{dx}
may be written in the form $\d \L = - (\p_\a \L \ \d x^\a
+ Q_\a^\b \  \p_\b  \d x^\a)$, from which will follow the
identity that captures the coordinate invariance
of $S_0^{\cal M}$.

Since $S_0^{\cal M}$ is unchanged by an arbitrary infinitesimal
coordinate transformation, we have that
\begin{eqnarray}
\label{currents}
0 = \d S_0^{\cal M} &= & -
\int_\M [Q^\b_\a \ \p_\b \d x^a + \p_\a\L \ \d x^\a]
+ \int_{\p \M} {\cal L} n_\a \d x^\a \cr
&=& - \int_\M \p_\b [\L \d^\b_\a - Q^\b_\a ] \ \d x^\a
+ \int_{\p \M}   [\L \d^\b_\a - Q^\b_\a ] n_b \d x^\a
\end{eqnarray}
where $n_\b$ is the outward pointing normal vector field to $\p \M$.
Thus, we may conclude that $\p_\b[\L \d^\b_\a - Q^\b_\a]$
vanishes in the interior and that
$n_\b[\L \d{}^\b_\a - Q^\b_\a]$ vanishes on the boundary.
Since the fields themselves are unrestricted on $\p \M$ (only
the variations of the fields are constrained), we must in fact
have
\begin{equation}
\label{cc}
\L \d^\b_\a - Q^\b_a = 0
\end{equation}
{\it identically} everywhere.  This is just the generalization of the
familiar statement that, on a $0+1$ dimensional spacetime, the
Hamiltonian constructed from a diffeomorphism invariant
action for a system of scalar fields vanishes identically.

The result \ref{cc} may be used to show that the action \ref{newS}
provides an acceptable variational principle when
$S^{\cal M}$ is varied within a class of
histories for which the fields are fixed on
the $f=0$ surface.  Specifically, fix some $n-1$
manifold $\Sigma$ and an embedding $\eta: \Sigma \rightarrow \M$ such that
$\M - \Sigma$ has exactly two connected components (which we arbitrarily
call the inside and the outside).  Note that $\Sigma$ may have
a boundary $\partial \Sigma$.
Consider the class of histories for which the surface defined by
$f=0$ gives the above embedding of $\Sigma$ into $\M$ up to
a diffeomorphism and for which $f>0$ inside and $f<0$ outside.  The
inside may still contain part of $\p \M$, although the
most interesting case is when $\p \M$ lies completely outside.  Furthermore,
we will vary the histories only within the class for which all type
I fields are fixed (up to a diffeomorphism of $\M$)
on the part of $\p \M$ inside of $f=0$
and on the $f=0$ surface.  This may be done, for example, by
choosing histories for which some set of scalar fields may be
used as a coordinate system near $\p \M$ and $f=0$ and for which the
components of the tensor fields have some fixed relationship with
these scalars and their gradients.

A direct variation of $S_0$, together with the usual integrations
by parts yields:

\begin{eqnarray}
\label{EL}
\d S^{\cal M} &= \int_\M \Biggl( \theta(f) &\Bigl[ \
\Bigl( {{\p \L} \over {\p  \f}} -
\D_\r {{\p \L} \over {\p (\D_\r \f)}} \Bigr) \d \f \cr
&& + \Bigl( j \f \d^\b_\a + \sum_k \fd \d^\b_{\mu_k}
- \sum_k \fu \d^{\nu_k}_\a \Bigr) {{\p \L} \over {\p (\D_\r \f)}}
\delta \Gamma^\a_{\r \b } \Bigr]  \cr
&& + \delta(f) \Bigl[ \d f \L - {{\p \L} \over {\D_\r \f}} \p_\r f \
\delta \f \Bigr]
\Biggr) \cr
&+& \int_{\p \M} \theta(f) n_\r {{\p \L} \over {\p (\D_\r \f)}} \delta \f
\end{eqnarray}
where $\delta(f)$ is the Dirac delta-function and is not to be
confused with the variation $\delta f$.
The first line in \ref{EL} contains just the usual Euler-Lagrange
equations inside the surface $f=0$ and the second line explicitly
displays the conspiracies that make $\L$ independent of the background
connection.  Line four contains the usual boundary terms on the
inside part of $\p \M$, which vanish for our class of variations.
The third line contains the terms of interest; in order
that our new variational principle not restrict the dynamics excessively,
we must show that these terms vanish when the variations satisfy the
boundary conditions given above and when the Euler-Lagrange equations
are satisfied on the $f=0$ surface\footnote{As may be seen from
a brief study of the
free relativistic particle, an attempt to leave these variations
arbitrary (as in a naive application of \ref{newS})
and use their coefficients as additional equations of
motion leads to nonsense.}.

To do so, note that the variation $\delta f$ will move the $f=0$
boundary surface.  Since all fields, including $f$ itself, are
specified on the $f=0$ surface up to diffeomorphism,
the variations $\delta \f$ must be just those
induced by some diffeomorphism
$x^\a \rightarrow x^\a +\delta x^\a$, which were explicitly
displayed in equation \ref{dx}.  Similarly, $\delta f =
- \p_\a \delta x^\a$.   Thus these variations, when evaluated on the
$f=0$ surface, are not independent.  The relationships between the
$\delta \f$ will conspire, together with the general covariance of
$S_0^{\cal M}$, to make the term proportional to $\delta(f)$ vanish when the
usual Euler-Lagrange equations of $S_0^{\cal M}$ are imposed on the $f=0$
surface.

Now, a comparison of \ref{dx} and \ref{EL} shows that use of the Euler-Lagrange
equations for $\Gamma^{\sigma}_{\alpha \beta}$ greatly
simplifies the term proportional to $\delta(f)$ since, when
contracted with ${{\p \L} \over {\p (\D_\r \f)}}$, $\delta \f$
may be replaced by $\p_\a \f \ \delta x^\a$.  Thus, we find
\begin{equation}
\label{diff}
\delta S^{\cal M} - \theta(f) \delta S_0^{\cal M} = \int_\M \delta(f)
[\L \d^\b_\a - {{\p \L} \over {\p (\D_\b \f)}} \p_\a \f] \p_\b f
\delta x^\a
\end{equation}
which looks suspiciously like \ref{cc}.

Returning to the definition of $Q_\a^\b$ in \ref{dx},
the Euler-Lagrange equations for $\f$ show that we have
\begin{equation}
\label{almost}
Q^\b_\a = \D_\p \Bigl[ {{\p \L} \over {\p (\D_\r \f)}}
{{\p (\delta\f)} \over {\p (\p_\b \delta x^a)}}\Bigr]
+ {{\p \L} \over { \p (\D_\r \f)}} \D_\a \f \delta^\b_\r.
\end{equation}
where we have written the transformation $\delta \f$ displayed in
\ref{dx} as a function of $\d x^{\sigma}$ and $\p_\b \d x^\a$.
As before, the term in brackets vanishes by the Euler-Lagrange
equations for $\Gamma^{\sigma}_{\alpha \beta}$.  In the second
term, the contraction of these same Euler-Lagrange equations with
$\Gamma^{\sigma}_{\a \gamma}$ on the indices $\sigma$ and $\gamma$
shows that $\D_\r \f$ may be replaced with $\p_\r \f$ in \ref{almost}.
Equation \ref{cc} then reads
\begin{equation}
{\cal L} \delta^\b_\a - {{\p \L} \over {\p (\D_\r \f)}}
\p_\a \f \delta^\b_\r = 0
\end{equation}
so that $\delta S_0^{\cal M}$ vanishes when the
Euler-Lagrange equations hold inside and on the surface $f=0$.

Thus, the action \ref{newS} provides a perfectly valid variational
principle for our partial system.
This leads to the interesting question of how path integrals
based on \ref{newS} differ from those based on \ref{s0}.  Note that
in a canonical framework, and as opposed to the traditional approach
of \cite{Claudio,JJ}, use of an action of the form \ref{newS} allows the
lapse $N$ to be {\it completely} fixed along with the gauge freedom.
It seems likely that path integrals of both types are appropriate
to the study of diffeomorphism invariant systems,
though with different interpretations
which are yet to be fully understood.

\acknowledgments
The author would like to express his thanks to Steve Carlip,
Fay Dowker, Jim Hartle,
Gary Horowitz, and Jorma Louko for sharpening his thinking on this subject
and for suggesting important referecnes.
This work was supported by NSF grant PHY-908502.

\end{document}